\documentclass[amsmath,amssymb,aps,twocolumn,pra,prd,longbibliography]{revtex4-2}
\usepackage{amssymb}
\usepackage{graphicx}
\usepackage{dcolumn}
\usepackage{bm}
\usepackage{amsmath}
\usepackage{subfigure}
\usepackage{float}
\usepackage{color}
\usepackage[colorlinks=true,citecolor=blue]{hyperref}

\begin{document}


\title{Microwave spectroscopy and Zeeman effect of cesium $(n+2)D_{5/2}\rightarrow nF_{J}$ Rydberg transitions}

\author{Jingxu Bai$^{1}$}
\author{Rong Song$^{1}$}
\author{Zhenhua Li$^{1}$}
\author{Yuechun Jiao$^{1,2}$}
\thanks{ycjiao@sxu.edu.cn}
\author{Georg Raithel$^{3}$}
\author{Jianming Zhao$^{1,2}$}
\thanks{zhaojm@sxu.edu.cn}
\author{Suotang Jia$^{1,2}$}

\affiliation{
$^{1}$State Key Laboratory of Quantum Optics and Quantum Optics Devices, Institute of Laser Spectroscopy, Shanxi University, Taiyuan 030006, People's Republic of China\\
$^{2}$Collaborative Innovation Center of Extreme Optics, Shanxi University, Taiyuan 030006, People's Republic of China\\
$^{3}$Department of Physics, University of Michigan, Ann Arbor, Michigan 48109-1120, USA}

\date{\today}

\begin{abstract}
We report on high-resolution microwave spectroscopy of cesium Rydberg $(n+2)D_{5/2}\rightarrow nF_{J}$ transitions in a cold atomic gas. Atoms laser-cooled and trapped in a magnetic-optical trap are prepared in the $D$ Rydberg state using a two-photon laser excitation scheme. A microwave field transmitted into the chamber with a microwave horn drives the Rydberg transitions, which are probed via state selective field ionization. Varying duration and power of the microwave pulse, we observe Fourier side-band spectra as well as damped, on-resonant Rabi oscillations with pulse areas up to $\gtrsim 3 \pi$. Furthermore, we investigate the Zeeman effect of the clearly resolved $nF_J$ fine-structure levels in fields up to 120~mG, where the transition into $nF_{7/2}$ displays a thee-peak Zeeman pattern, while $nF_{5/2}$ shows a two-peak pattern. Our theoretical models explain all observed spectral characteristics, showing good agreement with the experiment. Our measurements provide a pathway for the study of high-angular-momentum Rydberg states, initialization and coherent manipulation of such states, Rydberg-atom macrodimers, and other Rydberg-atom interactions. Furthermore, the presented methods are suitable for calibration of microwave radiation as well as for nulling and calibration of DC magnetic fields in experimental chambers for cold atoms.
\end{abstract}


\maketitle


\section{Introduction}\label{Sec1}
Rydberg atoms with principal quantum numbers $n\gtrsim10$ have exaggerated properties~\cite{Gallagher}, such as long lifetimes ($\sim n^3$), large microwave electric-dipole transition matrix elements ($\sim n^2$), enormous electric polarizabilities ($\sim n^7$), and strong van der Waals~($\sim n^{11}$) and other Rydberg-atom interactions~\cite{Browaeys2020,Jiao2022}. These properties 
are the foundation for recent developments in single-atom manipulation~\cite{Browaeys2020,Srakaew2023}, single-photon sources~\cite{Petrosyan2018,Shi2022,Brito2021,Huerta2020}, quantum information and simulation~\cite{Nguyen2018,Jaksch2000,Urban2009}, as well as precise quantum  measurements~\cite{Jing2020,Sedlacek2012,Holloway2014,Hao2020,Bai2019,Gordon2014}. 
The coupling of nuclear, $\bf{I}$, and electronic angular momenta, $\bf{J} = \bf{L} + \bf{S}$, causes hyperfine and fine structure splittings and shifts~\cite{Corney2006,Steck} [$\bf{L}$ and $\bf{S}$ denote orbital and spin angular momentum of the electron(s)].
The high sensitivity of Rydberg atoms to 
microwave fields, combined with their Zeeman and Paschen-Back effects, leads to rich spectroscopic structures. In room temperature cells, Zhang $et~al.$~\cite{Zhang2018,Bao2016} have observed the Zeeman effect of Cs Rydberg atoms in a magnetic field. Cheng $et~al.$~\cite{Cheng2017} have conducted high-precision spectroscopy of Rydberg-level splittings in an external magnetic field with two different optical polarization combinations of $\sigma^+\pm\sigma^-$ for probe and coupling laser beams in ladder-type electromagnetically induced transparency (EIT). 
In a cesium cold-atom experiment with a controllable magnetic field, Xue $et~al.$~\cite{Xue2019} have investigated the influence of Zeeman splittings, optical pumping, and probe-light-induced radiation pressure
in cold-atom Rydberg-EIT spectra. Recently, laser frequencies could be stabilized to Rydberg transitions using Zeeman-modulation Rydberg-EIT spectroscopy~\cite{Jia20201,Jia20202}. This method is applicable to a variety of applications in modern-physics research, including quantum information processing~\cite{Jaksch2000,Urban2009}, many-body dynamics~\cite{Ma2017}, and Rydberg molecules~\cite{Bottcher2016}. 

Microwave spectroscopy of Rydberg atoms, which typically has a resolution limit in the range of about 10~kHz,
is well-suited to measure energy levels and quantum defects of Rydberg states for which no convenient laser excitation schemes exist, such as high-$\ell$ ($\ell$=3,4,and 5) states of rubidium~\cite{Berl2020,Lee2016,Moore2020,Li2003} and cesium~\cite{Deiglmayr2013,Goy,bai2023}. Recently, Cardman $et~al.$~\cite{Cardman2022} have measured the hyperfine structure (HFS) and Zeeman effect of $nP_{1/2}$ Rydberg states using narrow-linewidth mm-wave spectroscopy in a cold $^{85}$Rb ensemble.

In the present work, we report on high-resolution microwave spectroscopy of $(n+2)D_{5/2}\rightarrow nF_{J}$ transitions by scanning a weak microwave field in a cold Cs Rydberg atom sample. In our first study we change the microwave pulse duration and power at zero magnetic field to establish the spectral resolution. We observe symmetric, high-contrast Fourier-sideband spectra that demonstrate coherent excitation for microwave pulses of up to several microseconds in duration, while at $20~\mu$s we reach our linewidth limit of $\sim$140~kHz. The fine structure (FS) splitting of the $nF_J$ levels is clearly resolved in all cases.
In the second part of our work, we apply a controlled homogeneous magnetic field using a pair of Helmholtz coils, and perform high-precision Zeeman spectroscopy for both FS components $nF_{5/2}$ and $nF_{7/2}$.
We observe two-peak and three-peak Zeeman splitting patterns, respectively, in agreement with calculations. The high-resolution Zeeman spectroscopy of high angular momentum ($\ell$=3) Rydberg states, presented in our work, will be helpful for in-situ calibration of magnetic fields in cold-atom and cold-molecule research, it may enhance the accuracy of future studies of high-$\ell$ quantum defects, and it may serve well in future quantum-control and -simulation studies involving high-$\ell$ Rydberg states.

\section{Experimental setup}\label{Sec2}

\begin{figure}[htbp]
\begin{center}
\includegraphics[width=0.45\textwidth]{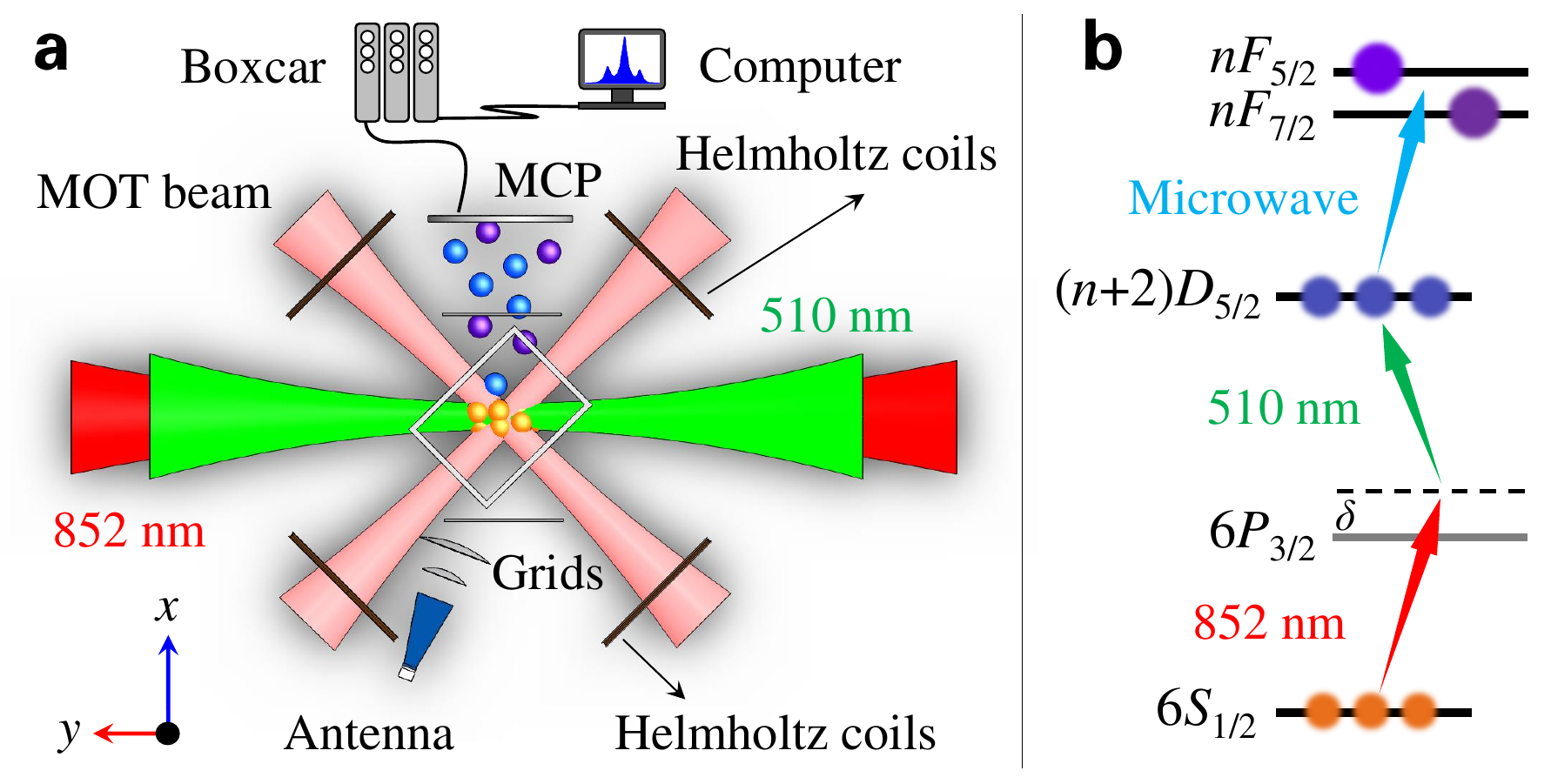}
\end{center}
\caption{(color online). (a) Schematic of the experimental setup. Cesium atoms are cooled and trapped at the center of a MOT. Two excitation lasers (wavelengths 852~nm and 510~nm) are overlapped and counter-propagated through the MOT to prepare $(n+2)D_{5/2}$ Rydberg  atoms. A tunable microwave field, applied via a microwave horn, allows us to perform high-resolution spectroscopy of $(n+2)D_{5/2} \to nF_J$ transitions. Three pairs of orthogonal electric grids and Helmholtz coils are employed to compensate stray electric and magnetic fields, respectively, and to scan the magnetic field in $z$-direction for Zeeman-effect studies. Rydberg-atom electric-field ionization and time-resolved ion detection allow us to measure Rydberg-transition probabilities. See text for details. (b) Energy level diagram. Two-photon excitation using the 852~nm and 510~nm lasers (respective single-photon Rabi frequencies $\Omega_{852}$ and $\Omega_{510}$ on the $6S_{1/2}\rightarrow 6P_{3/2}$ and $6P_{3/2}\rightarrow (n+2)D_{5/2}$ transitions, and intermediate-state detuning $\delta=330$~MHz) yields $(n+2)D_{5/2}$ Rydberg atoms. The tunable microwave field drives $(n+2)D_{5/2}$ to $nF_{J}$ transitions.
}\label{Fig1}
\end{figure}

The experimental setup is shown in Fig.~\ref{Fig1}(a). Cesium atoms are cooled and trapped in a magneto-optical trap (MOT) with a density of $\approx 10^{10}~\mathrm{cm}^{-3}$ and a temperature of $\approx 100~\mu$K. After switching off the MOT and waiting for a delay time of 1~ms, we apply the 852-nm and 510-nm Rydberg-excitation laser pulses, which have respective
Gaussian beam-waist parameters of $\omega_{852}$ = 750~$\mu$m and $\omega_{510}$ = 1000~$\mu$m. The two-photon optical Rydberg-excitation Rabi frequency in units Hz, at the beam centers, is kept at $\Omega_{2Ph}=\Omega_{852}\Omega_{510}/(2\delta) \approx 9$~kHz, which for our MOT atom density and optical excitation pulse length of $t=1~\mu$s results in a Rydberg-atom excitation probability of $(2\pi\Omega_{2Ph} \, t)^2/4 \approx 0.8 \times10^{-3}$. This is sufficiently low that Rydberg-atom interactions play no predominant role in our experiment. The microwave field ranges in frequency from 22~GHz to 33~GHz and is emitted by a small antenna into the chamber. Since the chamber is largely made from metal, the calibration factor between microwave field strength and square-root of power must be determined experimentally based upon the measured atomic response (see Sec.~\ref{subA}).

Three pairs of electric grids with symmetry points centered at the MOT are placed on the $x$-, $y$- and $z$-axes (in Fig.~\ref{Fig1} we only show the pair of grids on the $x$-axis). The grids are used to compensate stray electric fields and to apply an electric-field ramp for state-selective electric-field ionization and detection of the Rydberg atoms. After field compensation, the stray electric field during laser and microwave excitation is reduced less than 2~mV/cm~\cite{bai2023}. The liberated ions are accelerated to reach the MCP detector. Time-resolved ion signals, recorded with a boxcar integrator and computer, yield the Rydberg-state populations. Furthermore, three orthogonal pairs of Helmholtz coils are placed around the chamber to compensate stray magnetic fields and to apply controlled magnetic fields to investigate the Zeeman splitting of the $(n+2)D_{5/2}$ to $nF_{J}$ microwave transitions. The stray magnetic field is less than 5~mG after compensation~\cite{bai2023}.

Fig.~\ref{Fig1}(b) shows the level diagram.
The $(n+2)D_{5/2}$ state is populated via
two-photon laser excitation from $6S_{1/2}$ into $(n+2)D_{5/2}$.
The two-photon transition is
blue-detuned from the intermediate $6P_{3/2}$ state
by $\delta = 330$~MHz in order to reduce photon scattering and light radiation pressure.
In the experiment, the 852-nm laser is locked with modulation-free polarization spectroscopy~\cite{Pearman2002}, and the 510-nm laser is locked to a Fabry-Perot cavity with a finesse of $15\times 10^3$  using the Pound-Drever-Hall technique~\cite{Black2001}. The linewidths of both locked lasers are less than 100~kHz. The frequency of the microwave field, which is radiated into the chamber from a microwave horn, is scanned
across selected $(n+2)D_{5/2}$ to $nF_{J}$ transitions
to obtain narrow-linewidth Rydberg microwave spectra.

\section{Results and discussions}\label{Sec3}

\subsection{Fourier sideband spectroscopy}\label{subA}

\begin{figure}[htbp]
\begin{center}
\includegraphics[width=0.45\textwidth]{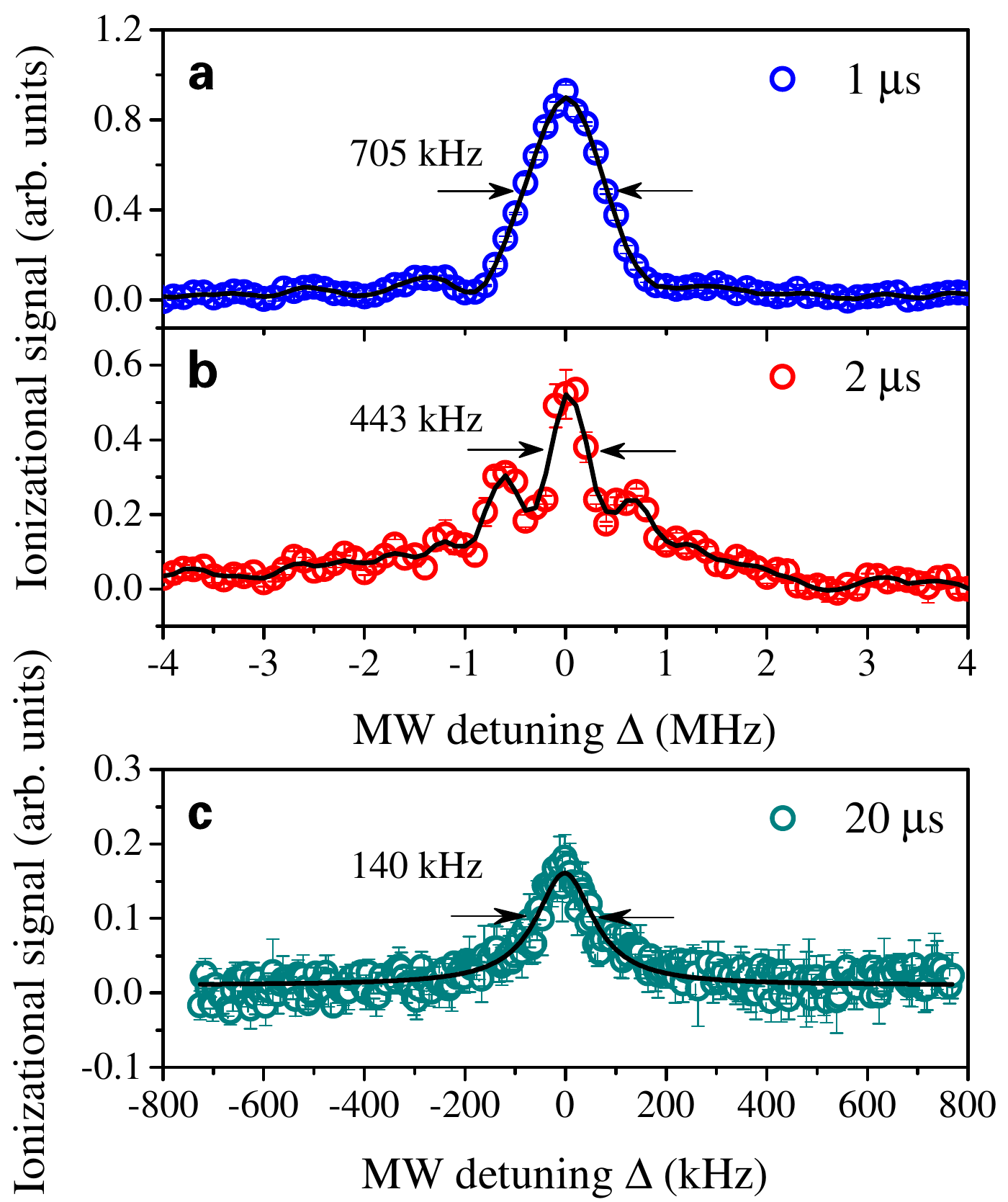}
\end{center}
\caption{(color online). Measurements of microwave spectra of the $48D_{5/2}$ $\to$ $46F_{7/2}$ transition with the power of the signal generator set to $-50$~dBm, a scan step size 100~kHz, and pulse durations of (a) 1~$\mu$s and (b) 2~$\mu$s. Black lines show smoothed curves obtained with the Savitzky-Golay  method. We clearly observe Fourier sidebands. The spectrum in (a) is below saturation, while the one in (b) is somewhat saturated (details see text).
(c) Spectrum at $-72$~dBm, scan step size 5~kHz, and pulse duration $\tau=20~\mu$s. The Lorentzian fit (black curve) has a full width at half maximum (FWHM) of 140~kHz. Error bars show the standard error of the mean (SEM) of the samples.} \label{Fig2}
\end{figure}

In our first experiment, the frequency of the 510-nm laser is tuned to excite the $48D_{5/2}$ state. Electric and magnetic fields are set to zero, within the uncertainties of the field compensation routine.
The microwave field then drives the transition $48D_{5/2}$ $\to$ $46F_J$. Microwave spectra are obtained by state-selective field ionization and gated ion detection, as the microwave frequency is scanned across the $48D_{5/2}$ $\to$ $46F_{7/2}$ transition
(center frequency 29.74323~GHz and scan ranges from 1.6~MHz to 8~MHz).
In Fig.~\ref{Fig2}(a), we present a typical microwave spectrum of the transition for a microwave pulse duration of $\tau$=1~$\mu$s. The spectrum in Fig.~\ref{Fig2}(a) has a FWHM linewidth of 705~kHz. This approximately accords with the Fourier limit of 890~kHz for
the case of non-saturated excitation, where the excitation spectrum is the square of a sinc-function with a FWHM of $0.89/\tau$.

In order to reduce the linewidth, in Figs.~\ref{Fig2}(b) and (c)
we increase the pulse duration $\tau$ to $2~\mu$s and 20~$\mu$s, respectively. The width of the central lobe in Fig.~\ref{Fig2}(b) agrees well with the (unsaturated) Fourier limit of 445~kHz; however, the overall width of the spectrum, including its side lobes, is considerably wider due to saturation. The experimental resolution limit of 140~kHz is reached in Fig.~\ref{Fig2}(c), where we have reduced the Rabi frequency by about a factor of 12.5 and increased $\tau$ by a factor of 10 relative to Fig.~\ref{Fig2}(b). The FWHM width in Fig.~\ref{Fig2}(c) is about three times larger than the Fourier limit (which is 45~kHz for a $\tau$= 20~$\mu$s pulse). The excess width is attributed to dipolar Rydberg-atom interactions and residual Zeeman shifts of the different sublevels in the residual stray magnetic field (for more details see our previous work~\cite{bai2023}).

\begin{figure*}[htbp]
\begin{center}
\includegraphics[width=0.95\textwidth]{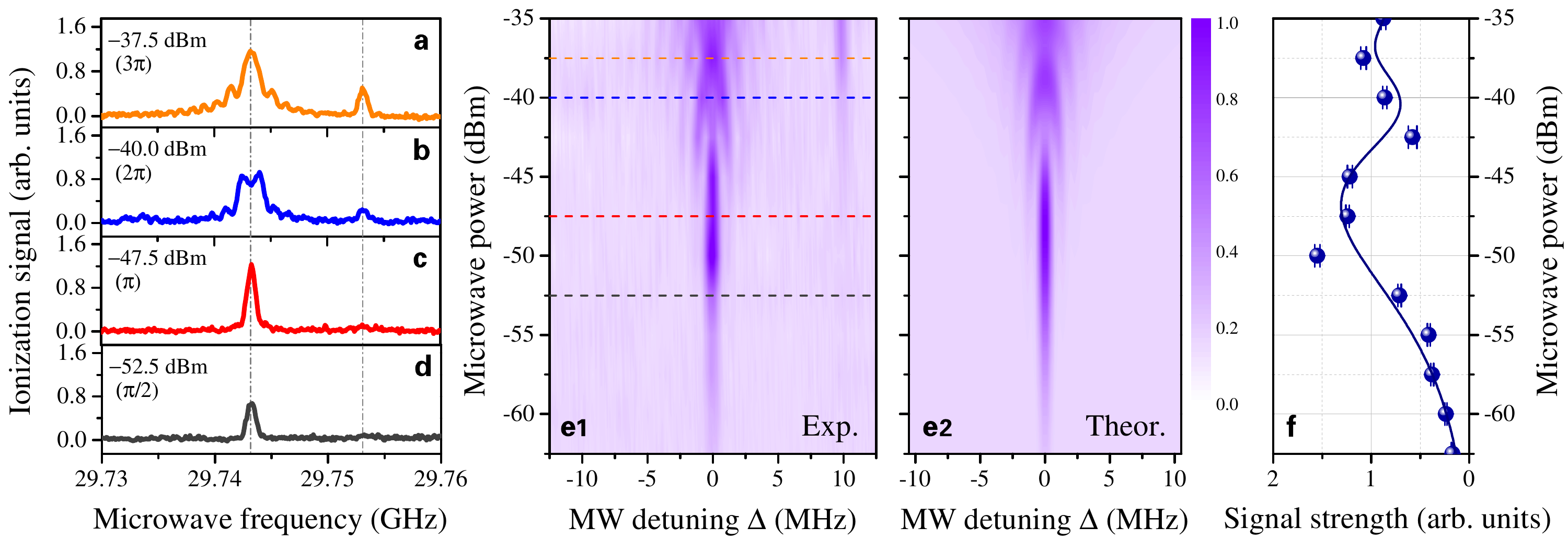}
\end{center}
\caption{(color online) Measurements of microwave spectra of $48D_{5/2}\rightarrow46F_{J}$ transition as a function of microwave frequency for the indicated microwave powers and estimated pulse areas for resonant excitation of $46F_{7/2}$ and microwave pulse duration of 1~$\mu$s. The two vertical gray dash-dotted lines at 29.743~GHz and 29.753~GHz mark the resonant frequencies for  $46F_{7/2}$ (left) and $46F_{5/2}$ (right), respectively. Each spectrum is an average over 20 measurements.
(e1) Contour plot of microwave spectra as a function of microwave detuning $\Delta$ and for the power ranging from $-62.5~\mathrm{dBm}$ to $-35~\mathrm{dBm}$ in steps of 2.5~dBm. The four horizontal dashed lines mark the spectra of (a) (orange), (b) (blue), (c) (red) and (d) (gray), respectively. (e2) Simulation result for (e1). (f) Measured signal strength of the resonant transition $48D_{5/2} \to 46F_{7/2}$ as a function of microwave power, showing a damped Rabi oscillation, and fit (see text).}\label{Fig3}
\end{figure*}

Next we comment on the Fourier-sideband strengths. In the case of weak saturation, where the microwave Rabi frequency $\Omega < |\Delta|$ on the Fourier sidebands, the amplitude of the sidebands rapidly diminishes with order (the first sidebands have a strength of 4.7\% of the central peak, the second sidebands have 1.6\%, etc). Fig.~\ref{Fig2}(a) is near the weak-saturation limit. With increasing saturation, where $|\Delta| \lesssim \Omega$ within at least the first sideband, the sidebands rise in strength relative to the central peak, as is evident in Fig.~\ref{Fig2}(b).

In order to exhibit the spectral dependence on both microwave power and $\Delta$, as well as to estimate the degree of coherence in the excitation process,
in Fig.~\ref{Fig3} we show a series of measured microwave spectra for microwave powers in the range between $-35$~dBm and $-62.5$~dBm and over a detuning range $|\Delta| \lesssim $15~MHz, for fixed $\tau =1~\mu$s. In Fig.~\ref{Fig3}(a-d), we present selected microwave spectra of $48D_{5/2}\rightarrow46F_{J}$ for microwave powers as indicated. Gray vertical dashed lines mark the resonant transition frequencies of 29.743~GHz and 29.753~GHz for $46F_{7/2}$ and $46F_{5/2}$, respectively. Due to the interaction between the atomic core and the valence electron, the FS of the Rydberg $F$-state is inverted, resulting in a greater transition frequency for $46F_{5/2}$ than for $46F_{7/2}$. Also, since
angular transition matrix elements for $46F_{7/2}$ are much greater than those for $48F_{5/2}$, the peak intensity of the $46F_{7/2}$ exceeds that of the $46F_{5/2}$ signal. The $46F_{7/2}$ line also is considerably more susceptible to broadening due to dipole-dipole interactions than the $46F_{5/2}$ line~\cite{bai2023}.

While the radial transition matrix elements for the $48D_{5/2}\rightarrow46F_{7/2}$ and the $48D_{5/2}\rightarrow46F_{5/2}$ transitions are nearly identical, the squares of the angular matrix elements averaged over magnetic quantum numbers differ by a factor of 20, equivalent to 13~dB in microwave power. As a result, the onset of the $F_{5/2}$ line in Fig.~\ref{Fig3} is shifted to higher power relative to that of the $F_{7/2}$ line. The $48D_{5/2}\rightarrow46F_{5/2}$ transition remains un-saturated throughout the investigated power range and only becomes visible at powers above $-47.5$~dBm, while the $48D_{5/2}\rightarrow46F_{7/2}$ transition becomes saturated. Due to its larger Rabi frequency, the $46F_{7/2}$ line develops visible Fourier side bands at powers above about $-50$~dBm and becomes saturation-broadened [see Figs.~\ref{Fig3}(a-e2)]. In their respective unsaturated regimes, both lines have a FWHM of about 800~kHz, which is at the Fourier limit for $\tau=1~\mu$s.

The cuts through the spectrum shown in Fig.~\ref{Fig3}(a-d) correspond with approximate resonant pulse areas of $\pi/2$, $\pi$, $2 \pi$ and $3 \pi$ for the $48D_{5/2}\rightarrow46F_{7/2}$ transition, averaged over $m_j$. At the resonant frequency of the  $48D_{5/2}\rightarrow46F_{7/2}$ transition we observe a damped Rabi oscillation, as is evident in Fig.~\ref{Fig3}(f).

In our model for the spectra in Figs.~\ref{Fig2} and~\ref{Fig3}, we use the function,
\begin{widetext}
\begin{equation}
\label{eq:th1}
\sum_{m_j}
A P(m_j) \frac{\Omega_j^2}{\Delta^2 + \Omega_j^2} \big[
1 - \exp(-2 \pi \tau \, \gamma \, \sqrt{\Delta^2 + \Omega_j^2}\, )
\cos ( 2 \pi \tau \, \sqrt{\Delta^2 + \Omega_j^2}) \big] ,
\end{equation}
\end{widetext}
where $\Delta$ is the microwave detuning in Hz, $A$ is an amplitude fit parameter, $P(m_j)$ are the probabilities for the $m_j$ sub-levels of the $nD_{5/2}$-state (which we set equal), $\Omega_j$ are the $m_j$-dependent microwave Rabi frequencies,
and $\gamma$ is a damping parameter. For linearly polarized microwaves and adopting the microwave electric-field direction as quantization axis, the microwave Rabi frequencies are
\begin{equation}
\label{eq:th2}
\Omega_j = \frac{e a_0 R_{(n+2)D_{5/2}}^{nF_j} w_{D_{5/2},m_j}^{F_J,m_j} E_{MW}}{h} ,
\end{equation}
in units Hz. There, $R$ is the radial matrix element, which equals 2564 in atomic units for both $J$-values for $n=46$, $w$ is the angular matrix element, and $E_{MW}$ is the microwave electric field. The $w$ for $F_{7/2}$ are 0.3499, 0.4518, 0.4949 for $|m_j|$= 5/2, 3/2 and 1/2, respectively.

We have modeled the spectra in Fig.~\ref{Fig3}(e) as $\Delta$- and power-dependent Rabi oscillations according to 
Eqs.~(\ref{eq:th1}) and~(\ref{eq:th2}). The exponential damping term has a form $\propto \exp(-\Phi \, \gamma)$, where $\Phi$ is the pulse area in rad/s, $2 \pi \tau \sqrt{\Omega_j^2 + \Delta^2}$. Good agreement is found for a damping constant $\gamma \approx 0.088~$s/rad, {\sl{i.e.}}, we are able to observe several Rabi cycles. The visibility reduction of the Rabi oscillation observed for $\Phi \gtrsim 2 \pi$ is, in part, due to the 40-$\%$ variation of the Rabi frequencies $\Omega_j$ as a function of $m_j$. The $m_j$-inhomogeneity could be addressed by optical pumping of the sample and by using proper optical and microwave polarizations. The level of coherence in Fig.~\ref{Fig3} would suffice for applications in quantum control and quantum simulations using ensembles of $\ell=2$ and $\ell=3$ Rydberg atoms.

A comparison of measured and simulated spectral maps in Fig.~\ref{Fig3}(e) allows us to calibrate the microwave electric field at 29.743~GHz. Averaging our calibrations performed for three pulse areas ($\pi$,  $2 \pi$ and  $3 \pi$), we find $E_{MW} = 10^{P_{MW}/20} \times 7.5~$V/m$~\pm 10\%$, where $P_{MW}$ is the microwave power in dBm. The uncertainty reflects experimental noise and half of our power step size of 2.5~dBm. Our simulation for the $46F_{7/2}$-spectrum, plotted for this calibration in Fig.~\ref{Fig3}(e2), agrees very well with the measurement, lending credence to the calibration. Similar calibrations can be performed for other transitions. It is noted that the calibration generally depends on the microwave frequency, because the mode pattern inside the metal vacuum chamber must be assumed to have a complex frequency dependence. However, 
the calibration is assumed to be constant over the relatively short frequency ranges used for each set of $(n+2)D_{5/2}\rightarrow nF_{J}$ transitions.

\begin{figure}[htbp]
\begin{center}
\includegraphics[width=0.45\textwidth]{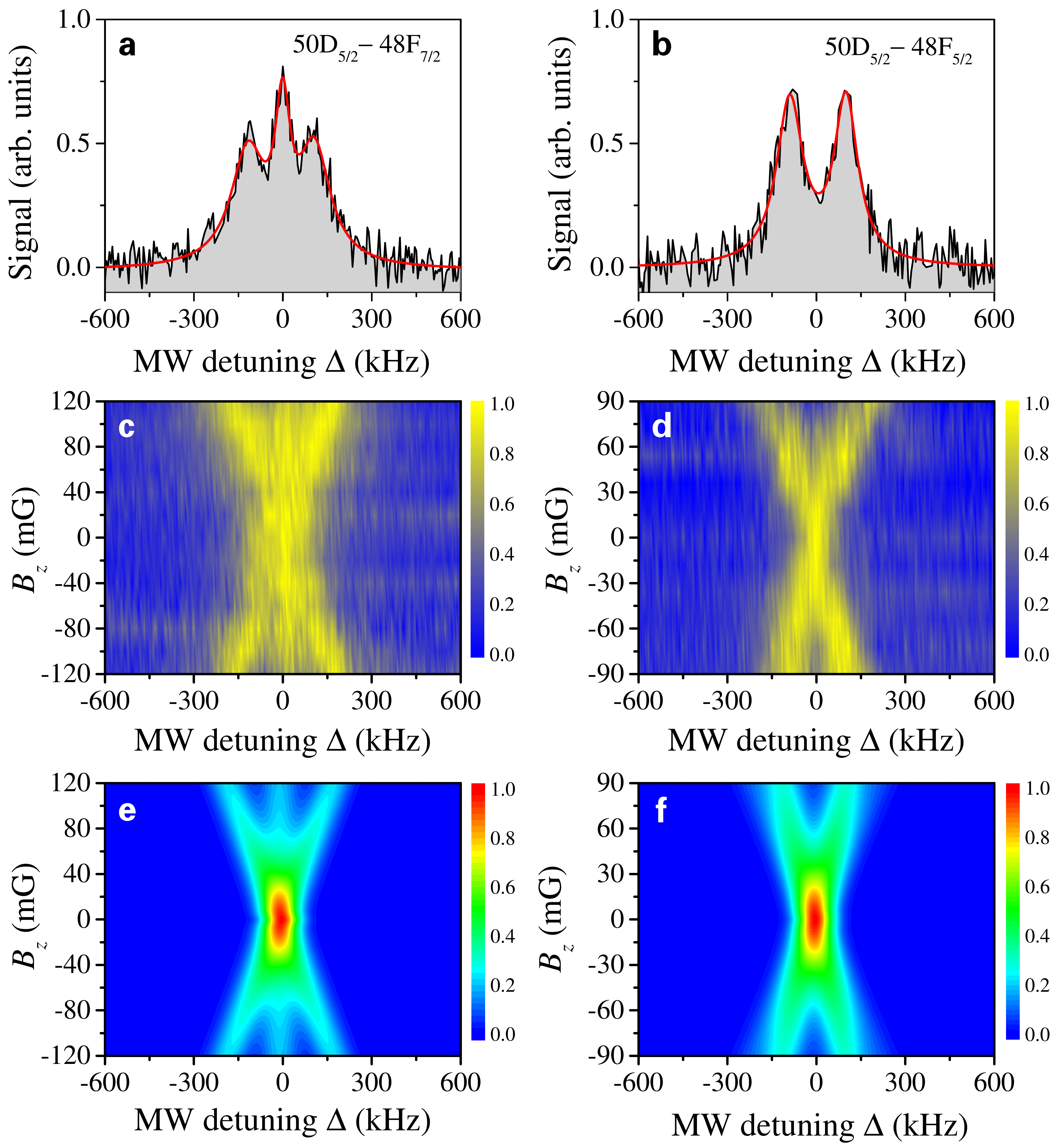}
\end{center}
\caption{(color online). Measurements and simulations of Zeeman spectra of the $50D_{5/2}\rightarrow 48F_{J}$ microwave transitions. In (a) and (b) we present Zeeman spectra for $J=7/2$ at $B=$83~mG and for $J=5/2$ at 58~mG, respectively. Red curves display the results of multi-peak Lorentzian fits. Contour plots of larger sets of measurement results are shown in (c) and (d) for $J=7/2$ and $J=5/2$, respectively. In panels (e) and (f) we present the results of corresponding simulations, in which the angle between the polarizations of the microwave and magnetic fields is set to $55^\circ$, which leads to good agreement between experiment and simulations. 
}\label{Fig4}
\end{figure}

\subsection{Zeeman spectroscopy}\label{subB}

In order to obtain narrow-linewidth microwave spectra in a weak magnetic field, we employ a microwave pulse duration $\tau$ = 20~$\mu$s in the Zeeman-splitting measurements presented in this section. After compensation of electric and magnetic stray fields, we apply a small current in the Helmholtz coils on the $z$-axis to generate an uniform magnetic field at the MOT center, which allows us to observe the Zeeman splitting of the $50D_{5/2} \to 48F_J$ transitions.

Figures~\ref{Fig4}(a) and (b) show Zeeman spectra for microwave fields driving the $50D_{5/2}\rightarrow 48F_{7/2}$  and $50D_{5/2}\rightarrow 48F_{5/2}$ transitions at magnetic fields of 83 and 58~mG, respectively. It is seen that the microwave spectra for $48F_{7/2}$ and $48F_{5/2}$ display different line profiles, namely three Zeeman peaks for $48F_{7/2}$ and two for $48F_{5/2}$. In the color-scale plots in Figs.~\ref{Fig4}(c) and (d) we present a wider set of measurements for $z$-magnetic fields covering ranges of $\pm 120~$~mG and $\pm 90$~mG for $48F_{7/2}$ and $48F_{5/2}$, respectively. It is confirmed that at fields exceeding several tens of mG the $48F_{7/2}$ and $48F_{5/2}$ spectra smoothly split into three and two lines, respectively.

\subsection{Theoretical analysis of Zeeman spectra}\label{subC}

\begin{figure}[htbp]
\begin{center}
\includegraphics[width=0.45\textwidth]{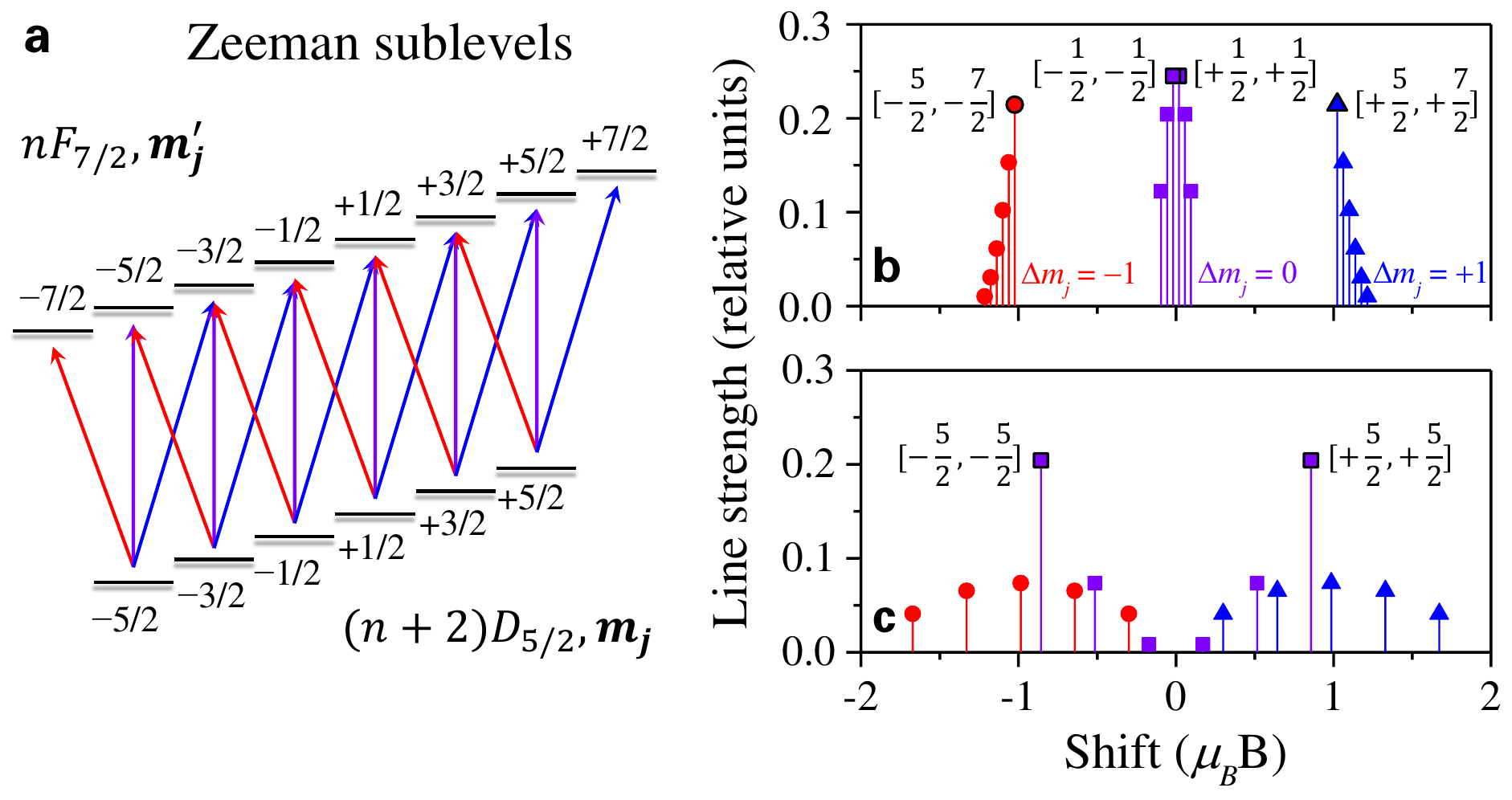}
\end{center}
\caption{(color online). (a) Zeeman sublevels for the  $(n+2)D_{5/2}\to nF_{7/2}$ transition in the presence of a weak magnetic field (main separation not to scale). Red, purple and blue arrows represent the transitions $\Delta m_j =-1$ ($\sigma^-$), $\Delta m_j =0$ ($\pi$) and $\Delta m_j =+1$ ($\sigma^+$), respectively. The anomalous Zeeman effect for $(n+2)D_{5/2}$ $\to$ $nF_{7/2}$ and  $(n+2)D_{5/2}$ $\to$ $nF_{5/2}$ produces the line patters shown in (b) and (c), respectively. For the line strengths we have assumed $x$- or $z$-polarized microwave electric fields; the former drives the $\sigma^{\pm 1}$-transitions with equal strengths, while the latter drives the $\pi$-transitions. For the $(n+2)D_{5/2}$ $\to$ $nF_{7/2}$ transition, the six transition shifts for each $\Delta m_j =0, \pm 1$  are tightly grouped and well separated, forming a three-peak Zeeman spectrum for cases of mixed microwave polarization. For the $(n+2)D_{5/2}$ $\to$ $nF_{5/2}$ transition, the six $\Delta m_j =0$ lines blend with the five-each $\Delta m_j =-1$ and $\Delta m_j =+1$ lines, generally forming a two-peak Zeeman spectrum, regardless of microwave polarization. The labels $[m_{j}$,$m'_{j}]$ mark the transitions with the largest transition probabilities.
}\label{Fig5}
\end{figure}

To understand the measured Zeeman spectra and the difference between the line profiles for the $nF_{7/2}$ and $nF_{5/2}$ upper states, we first plot all microwave-coupled Zeeman levels in Fig.~\ref{Fig5}(a) and calculate Zeeman shifts and transition matrix elements for all $m_j$ levels [ignoring the weak Rydberg HFS of the $(n+2)D_{5/2}$ lower state]. In a weak magnetic field, the $(n+2)D_{5/2}$ and $nF_J$ states are in the Zeeman regime of the FS, and the Zeeman shifts exhibit a linear dependence on the magnetic field. The weak magnetic field lifts the degeneracy of the $m_j$ sublevels, as shown in Fig.~\ref{Fig5}(a). According to electric-dipole selection rules, there are three transition types for changes in $m_j$, namely $\Delta m_j=-1,0,+1$, corresponding to $\sigma^-$, $\pi$ and $\sigma^+$ microwave polarizations. In Fig.~\ref{Fig5}(a), these are marked with red, purple and blue arrows, respectively.

Since in our work the magnetic field is in the Paschen-Back regime of the Rydberg HFS and the linear Zeeman regime of the FS, the level shifts from the field-free positions, visualized in Fig.~\ref{Fig5}(a) are,
\begin{equation}
\label{eq:th3}
\begin{split}
\Delta_{j}=\mu_B B (m_j g_j + g_I m_I) ,\\
\end{split}
\end{equation}
where $\mu_B$ is the Bohr magneton, $g_j$ and $g_I$ are the applicable electronic and nuclear $g$ factors, and $m_j$ and $m_I$ the corresponding magnetic quantum numbers. For our $(n+2)D_{5/2}$ and $nF_J$ Rydberg states of cesium the nuclear term is negligible. The Land\'{e} $g$-factors for the electronic Zeeman shifts of the relevant states,
\begin{equation}
\label{eq:th4}
g_j\simeq1+\frac{j(j+1)+s(s+1)-l(l+1)}{2j(j+1)} , 
\end{equation}
are $g_{D_{5/2}}$=1.2, $g_{F_{7/2}}$=1.1433, and $g_{F_{5/2}}$=0.8572. The transition shifts for $\Delta m_j=-1,0,+1$ can then be expressed as,
\begin{equation}
\label{eq:th5}
\begin{split}
\Delta(\sigma^-)&=\mu_B B \, [ m_j (g_{F_J} - g_{D_{5/2}}) - g_{F_J}]\\
\Delta(\pi)     &=\mu_B B \,   m_j (g_{F_J} - g_{D_{5/2}})\\
\Delta(\sigma^+)&=\mu_B B \, [ m_j (g_{F_J} - g_{D_{5/2}}) + g_{F_J}] .
\end{split}
\end{equation}

In Fig.~\ref{Fig5}(a) we plot the level scheme, and in Figs.~\ref{Fig5}(b) and (c) the relative transition strengths, $\vert w_{D_{5/2}, m_j}^{F_{J}, m'_j} \vert^2 $, versus the line positions from Eq.~(\ref{eq:th5}).
As seen in Fig.~\ref{Fig5}(a), for $(n+2)D_{5/2} \to nF_{7/2}$ there are six transitions for every $\Delta m_j$. Due to the small difference of $0.07$ between the lower- and upper-state $g_j$-factors, transitions with same $\Delta m_j$ cluster into tight groups, as is evident from Eq.~(\ref{eq:th3}) and seen in Fig.~\ref{Fig5}(b). For mixed microwave polarization, this leads to three peaks in the microwave spectra, in agreement with Figs.~\ref{Fig4}(a, c). The small difference in $g_j$-factors in combination with residual stray magnetic fields, as well as quite strong dipolar ``flip'' interactions between atom pairs in
$(n+2)D_{5/2}$ and $nF_{7/2}$~\cite{bai2023}, lead to line
broadening on the order of 100~kHz that keeps us from reaching the Fourier limit of the linewidth at zero $B$ field in Fig.~\ref{Fig4}(c). Notably, the three $\Delta m_j$-resolved lines at non-zero $B$-field are considerably narrower than the (single) line at $B$ near zero, pointing to dipolar interactions as the main cause for the line broadening at $B$ near zero.

Next, we consider the transition $(n+2)D_{5/2} \to nF_{5/2}$. In this case, the $g_J$-factors between the lower and upper levels differ by $0.343$, and the five $\Delta m_j = \pm 1$ and six $\Delta m_j = \pm 0$ transitions split by considerably more than in the transition $(n+2)D_{5/2} \to nF_{7/2}$, leading to a larger anomality of the Zeeman effect. Factoring in the distribution of oscillator strength, the overall line pattern features two lines, regardless of microwave polarization, and only the width of the two lines depends on the polarization. The insensitivity of the $(n+2)D_{5/2} \to nF_{5/2}$ Zeeman pattern to microwave polarization strongly differs from $(n+2)D_{5/2} \to nF_{7/2}$, where the Zeeman pattern strongly depends on microwave polarization. This behavior, which is evident from Figs.~\ref{Fig5}(b) and (c), is a result of the symmetry between line shifts and transition strengths, seen within the  $\Delta m_j = 0$ group as well as between the $\Delta m_j = \pm 1$ groups. In addition, for the case of $(n+2)D_{5/2} \to nF_{5/2}$ it also is important that $\pi$-transitions are strongest for large $\vert m_j \vert$, while the $\sigma^{\pm}$-transitions are strongest for small $\vert m_j \vert$ [see transition labels in Figs.~\ref{Fig5}(b) and (c)]. Further, the line broadening of the transition $(n+2)D_{5/2} \to nF_{5/2}$ near zero $B$-field is considerably less than that of the transition $(n+2)D_{5/2} \to nF_{7/2}$.
We attribute this difference to the relatively low strength of dipolar ``flip'' interactions between atom pairs in
$(n+2)D_{5/2}$ and $nF_{5/2}$~\cite{bai2023}.

For a comprehensive comparison between theory and measurement, we have simulated the Zeeman spectrum using the following principles.
(1) Ground-state $6S_{1/2}, m_j$  populations are isotropic. (2) Initial Rydberg-state $m_j$-populations in $(n+2)D_{5/2}$ are proportional to the squares of two-photon matrix elements for the $6S_{1/2}, m_{0,j}$ to $(n+2)D_{5/2}, m_j$ two-photon transitions for the given optical polarizations. This is justified by the small, far-from-saturation Rydberg excitation probabilities in our experiment. (3) Microwave Rydberg-transition probabilities are proportional to $\vert w_{D_{5/2}, m_j}^{F_{J}, m'_j} \vert^2 $. In Fig.~\ref{Fig4} this is justified because microwave powers were chosen well below saturating the Rydberg transitions. (4) All Zeeman Rydberg lines are convoluted with Gaussians of 100~kHz FWHM and are summed up to arrive at a net spectrum. (5) For the optical transitions, all HFS and FS effects are included. (6) For the microwave transitions, all HFS and FS effects are also included. However, the HFS of the $(n+2)D_{5/2}$ only has a minimal effect in the range of 10~kHz shifts, while the HFS of the $nF_{J}$ levels is entirely negligible. The simulated microwave Zeeman spectra are displayed in Figs.~\ref{Fig4}(e) and (f) for the $50D_{5/2}\rightarrow 48F_{7/2}$ and $50D_{5/2}\rightarrow 48F_{5/2}$ transitions, respectively. The simulation reproduces our experimental measurements well, including the two-peak profile for the $50D_{5/2}\rightarrow 48F_{5/2}$ transition and the three-peak profile for the $50D_{5/2}\rightarrow 48F_{7/2}$ transition. For the latter, it was necessary to assume an $\approx 55^\circ$ microwave polarization angle relative to the magnetic-field direction to achieve agreement. Such an angle is quite plausible due to the fact that the microwave is radiated into a complex metal chamber. Comparing measured data and simulated data for a range of polarization angles (not shown), the uncertainty of the polarization angle is determined to be about $\pm 5^\circ$. Further, there is a small shift in the simulation at zero magnetic field and near zero microwave detuning. This shift may be attributed to the HFS of $nD_{5/2}$, which is included in the simulation but is otherwise beyond the scope of our present work.

\section{Summary}\label{Sec4}

We have reported the observation of high-resolution microwave spectra of $(n+2)D_{5/2}\rightarrow nF_{J}$ transitions in cold cesium atom clouds.
For 20-$\mu$s long microwave pulses the observed spectral linewidth was 140~kHz,
while for 1-$\mu$s pulses coherent, Fourier-limited spectra with pulse areas up to about $3\pi$ were achieved, in agreement with a model that takes damping into account.
We have further studied the Zeeman effect of the fine-structure-split $(n+2)D_{5/2}\rightarrow nF_{J}$ transitions in fields up to about 100~mG. Results agree well with our theoretical models, which have provided detailed insight into the spectroscopic structure and the microwave electric-field strength and polarization within the atom-field interaction region.

In the future, Rydberg quantum simulators may benefit from our demonstrated coherent drive of relatively strong, high-$\ell$ transitions. Also, the method allows for microwave field-strength calibration and polarization estimation, which is useful for metal chambers in which computational approaches are prohibitive for field-mode estimation. Furthermore, the work is helpful to null and to calibrate DC magnetic fields in complex vacuum chambers utilizing optical methods and calibration-free atomic spectroscopy. Future simulation work for higher optical and microwave powers will likely require sophisticated methods to solve quantum master equations suitable to treat problems with many quantum states, such as quantum Monte Carlo wave-function simulations (QMCWF; see our previous work in~\cite{Xue2019}).  

\begin{acknowledgments}
This work is supported by the National Natural Science Foundation of China (Grants No. 61835007, 12120101004, 62175136, 12241408); the Scientific Cooperation Exchanges Project of Shanxi Province (Grant No. 202104041101015); the Changjiang Scholars and Innovative Research Team in Universities of the Ministry of Education of China (Grant No. IRT 17R70); and the 1331 project of Shanxi Province. G.R. acknowledges support by the University of Michigan.
\end{acknowledgments}

\bibliography{reference}

\end{document}